\documentstyle[prl,preprint,tighten,aps,epsfig]{revtex}
\begin{document}
\draft

\title{Quantum Superposition States of Bose-Einstein condensates}

\author{J. I. Cirac$^1$, M. Lewenstein$^2$, K. M\o lmer$^3$ and P. 
Zoller$^1$}
\address{(1) Institut f\"ur Theoretische Physik, Universit\"at Innsbruck 
\\
Technikerstrasse 25, A--6020 Innsbruck, Austria}

\address{(2) Commissariat a l'Enegie Atomique, DSM/DRECAM/SPAM\\  
Centre d'Etudes de Saclay, 91191 Gif-sur-Yvette, France}

\address{(3) Institute of Physics and Astronomy, University of
Aarhus, DK--8000 {\AA}rhus C}

\maketitle

\begin{abstract}
We  propose a scheme to
create a macroscopic ``Sch\"odinger cat'' state formed by
two interacting Bose condensates. In analogy with {\it quantum optics}, 
where
the control and engineering of quantum states can be maintained to a 
large extend, we consider the present scheme to be an example of {\it quantum 
atom optics} at work.

\end{abstract}


\narrowtext

\section{Introduction}

The recent experimental realization of  Bose-Einstein condensation  of
trapped cold rubidium \cite{BEC-Rb},  sodium
\cite{Ketterle}, and lithium \cite{BEC-Li} atoms has initiated new areas 
of atomic, molecular and optical physics \cite{BEC}. While some of these 
new areas remain still somewhat speculative, others have 
already attained  firm experimental grounds, and many
of them are based on the analogy between the matter waves
and electromagnetic waves, or between bosonic atoms and photons.

On the level of single atoms, the analogy between the matter and
electromagnetic waves has led to  rapid developments of {\it atom 
optics}
\cite{atop}. Some authors have thus considered
a possibility of {\it nonlinear atom optics}
in systems of many cold atoms, where the quantum statistical properties
and atom-atom interactions become important \cite{nao}. It has been also 
pointed out \cite{edwards} that nonlinear excitations of Bose-Einstein 
condensates (BEC) may lead to various analogs of nonlinear optics. Most 
of these theories have a mean field character, i.e. they disregard 
quantum fluctuations of the atomic field operators and
concentrate on the nonlinear Sch\"odinger-like wave 
equations for 
the matter
wave functions.

In many situations, such as for example laser cooling or
optical imaging, cold atoms not only exhibit their quantum statistical
properties, but on top of that they interact with photons. This fact 
motivated
the developments of {\it quantum field theory of atoms and photons} 
\cite{review}. Although this theory accounts in principle for
quantum fluctuations of both atomic  and electromagnetic fields, the 
attention has so far been focused predominantly on the latter.

The atom-photon analogy has also  triggered  the studies in the area
of {\it physics of atom lasers} \cite{atomlas}. An atom laser, or a 
boser is a matter wave analog of an ordinary laser. Quite recently, the 
possibility
of employing BEC as a source of coherent matter waves has been 
demonstrated in the remarkable experiments of the MIT group \cite{kett}.

We propose here  to proceed with  this analogy and to
look for the implementation of further elements of quantum optics 
in {\it quantum atom optics}. 
In our opinion, one of the major domains of concern of modern quantum 
optics is preparation, engineering, control and detection of quantum 
states in various systems that involve light-matter interactions 
\cite{examples}. By analogy, quantum atom optics, in the sense proposed, concerns preparation, engineering,
control and detection of quantum states in atomic systems.
 Recent studies of excitations in
trapped BEC belong to this category, although so far 
rather limited kinds of
time dependent perturbations of
the trapping potential \cite{becexc} were used and
only few types of  
excitation
have been investigated. 
Walsworth and You 
 have proposed a
method of selective creation of quasi-particle excitations in trapped 
BEC \cite{ron}.  Their method, referred to as 
spatial magnetic resonance 
method, could in principle allow for engineering and control of arbitrary 
excitation
in the Bose condensed system.

One of the most spectacular achievements of quantum optics in the
recent years has been the observation and study of macroscopic 
(or strictly speaking mesoscopic) ``Schr\"odinger cat'' states of a 
trapped ion \cite{wineland}, and of  an
electromagnetic field in a high Q 
cavity \cite{brune}. Sch\"odinger
\cite{schr} has introduced his famous ``cat'' states in order to 
illustrate the fundamental problem of the correspondence between the 
micro-- and macro--worlds: the fact that quantum superposition states 
are never observed on the macro level. As postulated by von Neumann 
\cite{vonn}, this is due to irreversible reduction of  superposition 
states into statistical mixtures. Such reduction
occurs in any quantum measurement process and leaves the considered
system in a mixed state in  a ``preferred'' basis, determined by the 
measurement. Modern quantum measurement theory \cite{zurek} describes 
the reduction process in terms of quantum decoherence  due to 
interactions of the system with environment \cite{modmea}. Experimental 
realization of ``cat''
states requires thus typically sophisticated means to avoid the 
decoherence
effects \cite{wineland,brune}.

In this  paper we demonstrate that it is feasible to prepare, control 
and detect a ``Schr\"odinger cat'' formed by two interacting Bose 
condensates
of atoms in different internal states. Atom-atom interactions in our 
model are mediated through atom-atom collisions and a Josephson-like
 laser coupling that interchanges internal atomic states in a coherent 
manner.
The theory of such bi-condensates restricted to
 collisional interactions only has been recently discussed in the 
Thomas-Fermi approximation \cite{tfbi}, and beyond \cite{esry,Busch}. 
Amazingly, the simultaneous condensation of $^{87}$Rb atoms in two 
internal states (F=2,M=2) and (F=1,M=-1)
has been recently achieved  at JILA \cite{bicon}, using a 
combination of evaporative \cite{revket} and sympathetic \cite{ours} 
cooling.
As pointed out by  Julienne {\it et al.} \cite{julliene}, the 
simultaneous condensation was possible due to a very fortunate ratio of 
elastic/inelastic collision rates for Rb atoms. For the moment, the 
perspectives of extending the result of \cite{bicon} to other atomic 
species  are not promising.
Nevertheless, one could expect that various ways of modifying
atomic scattering  lengths will be realized 
\cite{kagan}, which will allow to control atomic collision processes in 
a desired way.
The above comments apply to the case of magnetic traps. Once it becomes
possible to achieve Bose--Einstein condenstaion in, for example, far--off
resonance traps, this will open other possibilities of trapping particles
with different internal levels. This will allow one to
meet in real atomic systems the conditions discussed below for the preparation of 
Schr\"odinger cat states.

The plan of the paper is  the following. 
In section II we present the quantum 
field theoretical model of two trapped condensates, and its simplified
two-mode caricature. The detailed analysis of the two-mode model is 
carried
out in section III. We show that in some circumstances the ground state 
of the system becomes  a ``Schr\"odinger cat'', 
and that the system can be 
 prepared in such a state by adiabatically changing 
the strength of the 
Josephson-like laser coupling.
Various approximate solutions are tested here in comparison with the 
exact numerical solution of the problem. In section IV approximate 
solutions
of the complete quantum field theoretical model are found. They display 
the same qualitative behavior as the one obtained for the two mode 
model. Finally, in section V we discuss  the  
feasibility of experimental 
observation of ``Schr\"odinger cat'' states of two interacting Bose
condensates.

\section{Quantum field theory of two interacting condensates}

We consider here Bose--Einstein condensation of a trapped gas of atoms
with
two internal levels $|A\rangle$ and $|B\rangle$. The atoms interact via 
$AA$, $BB$, and $AB$ elastic collisions. Additionally, a set of laser 
fields drives
coherently a  Raman transition connecting $|A\rangle
\leftrightarrow |B\rangle$.
In the formalism of second quantization, such a system is described by 
the
following Hamiltonian
\begin{equation}
\label{Ham1}
H=H_A+H_B+H_{\rm Int}+H_{\rm Las}
\end{equation}
where
\begin{mathletters}
\label{Ham2}
\begin{eqnarray}
H_{A,B} &=& \int d^3\vec x \; \hat\Psi_{A,B}(\vec x)^\dagger
\left[ -\frac{\hbar^2}{2M}\nabla^2+\frac{1}{2}M\omega_{A,B}^2 x^2
+ \frac{4\pi \hbar^2 a^{\rm sc}_{A,B}}{2M}
\hat\Psi_{A,B}(\vec x)^\dagger \hat\Psi_{A,B}(\vec x) \right]
\hat\Psi_{A,B}(\vec x),\\
H_{\rm Int} &=& \frac{4\pi \hbar^2 a^{\rm sc}_{AB}}{M}
\int d^3\vec x \;\hat\Psi_A(\vec x)^\dagger
\hat\Psi_B(\vec x)^\dagger \hat\Psi_B(\vec x) \hat\Psi_A(\vec x)\\
H_{\rm Las} &=& - \frac{\hbar\Omega}{2} \int d^3\vec x
\left[ \hat\Psi_B(\vec x)^\dagger \hat\Psi_A(\vec x)e^{-i\Delta t}+
\hat\Psi_A(\vec x)^\dagger \hat\Psi_B(\vec x) e^{+i\Delta t}\right].
\end{eqnarray}
\end{mathletters}
Here, $H_{A,B}$ describes the evolution of atoms in $|A\rangle$
and $|B\rangle$, respectively, in the absence of interactions between
atoms in different internal states; $H_{\rm Int}$ describes the
interactions between atoms in $|A\rangle$ and $|B\rangle$ due to
collisions; $H_{\rm Las}$ describes the Raman transitions induced by  
the laser
detuned by $\Delta$ from the Raman resonance; such interactions act as a 
Josephson-like coupling
which transfers coherently particles  between $|A\rangle$ and 
$|B\rangle$, at  a Rabi frequency $\Omega>0$. Atoms are confined in
harmonic potentials of frequencies $\omega_{A,B}$, and $a^{\rm
sc}_{A,B,AB}$ are the scattering lengths for the corresponding
collisions, respectively.
We assume that the collisions are purely elastic, and that
they do not change the number
of particles in each internal level.

The field operators
$\hat\Psi_{A,B}(\vec x)$, $\hat\Psi_{A,B}(\vec x)^{\dagger}$ annihilate 
and create atoms at $\vec x$ in the internal state $|A\rangle$
and $|B\rangle$. They fulfill the standard bosonic 
commutation
relations
\begin{mathletters}
\label{comutatores}
\begin{eqnarray}
&[&\hat\Psi_{A}(\vec x),\hat\Psi_{A}(\vec x')^\dagger] =\delta
(\vec x-\vec x'),\\
&[&\hat\Psi_{B}(\vec x),\hat\Psi_{B}(\vec x')^\dagger] =\delta
(\vec x-\vec x').
\end{eqnarray}
\end{mathletters}
For the sake of simplicity, throughout this paper we will
assume that $a^{\rm sc}_{B}=a^{\rm sc}_{A}\equiv a^{\rm sc}$,
resonant laser excitation $\Delta=0$,
 and 
$\omega_B=\omega_A\equiv \omega$. This makes the Hamiltonian
(\ref{Ham1})  invariant under the exchange $A \leftrightarrow B$,
which simplifies  the analytical arguments. In experiment
(c.f.\cite{bicon})  this symmetric situation is not directly realized,
since atoms in different (F,M) states
 experience different Zeeman effects in the magnetic field, and feel thus 
different
trap potentials. Nevertheless, we stress that our assumption has only a 
technical
character. 
All the results presented below for the $A-B$ symmetric case can be 
translated
into the asymmetric case, as we shall see below.
In general, if $\omega_A\ne\omega_B$, one can always choose 
the detuning $\Delta\ne 0$ to compensate for the potential difference. 
One should also mention the fact that
the different Zeeman effects, combined with gravity, may  displace
the traps with respect to each other; even for such a sitation compensation
of potential difference using $\Delta\ne 0$ should be posible, although more
complex.

Let us now rescale the
variables to dimensionless ones as follows: First, we divide $H$ by
$\hbar\omega$, and then define $\lambda\equiv\Omega/(2\omega)>0$, $\vec
r\equiv\vec x/x_0$,
$\hat\Psi(\vec r)\equiv \hat\Psi(\vec x) x_0^{3/2}$, $U_0\equiv 4\pi 
a^{\rm sc}/x_0$,
and $U_1\equiv 4\pi a^{\rm sc}_{AB}/x_0$, where 
$x_0\equiv(\hbar/M\omega)^{(1/2)}$.
The rescaled dimensionless Hamiltonians (\ref{Ham2}) read now
\begin{mathletters}
\label{Ham3}
\begin{eqnarray}
H_{A,B} &=& \int d^3\vec r \; \hat\Psi_{A,B}(\vec r)^\dagger
\left[ -\frac{\nabla^2}{2}+\frac{r^2}{2}
+ \frac{U_0}{2}  \hat\Psi_{A,B}(\vec r)^\dagger  \hat\Psi_{A,B}(\vec r) 
\right]
 \hat\Psi_{A,B}(\vec r),\\
H_{\rm Int} &=& U_1 \int d^3\vec r \; \hat\Psi_A(\vec r)^\dagger
 \hat\Psi_B(\vec r)^\dagger  \hat\Psi_B(\vec r)  \hat\Psi_A(\vec r)\\
H_{\rm Las} &=& - \lambda \int d^3\vec r
\left[  \hat\Psi_B(\vec r)^\dagger  \hat\Psi_A(\vec r) +
 \hat\Psi_A(\vec r)^\dagger  \hat\Psi_B(\vec r) \right],
\end{eqnarray}
\end{mathletters}

Our objective  is to study the properties of the system described by
the above Hamiltonians at zero temperature. To this aim we need to find
the ground state of Hamiltonian (\ref{Ham1}) with
the corresponding terms defined in Eqs. (\ref{Ham3}). The 
search of this state is a
very difficult task. Under some
conditions one can obtain mean field
approximations, and numerical (approximate) results for the ground state 
of (\ref{Ham1}).  
In order  to understand better our model, we will first analyze a very simple two-mode model
described by  a caricature of the Hamiltonian (\ref{Ham1}). As we shall 
see in section III and IV, the analysis of the simplified model
resembles very much the analysis of the complete model described by 
(\ref{Ham1}).

The two-mode approximation of the Hamiltonian (\ref{Ham1}) is given by
(\ref{Ham1}) but with
\begin{mathletters}
\label{Ham4}
\begin{eqnarray}
H_{A} &=& \omega_A a^\dagger a + \frac{U_{AA}}{2} a^\dagger a^\dagger 
a a,\\
H_{B} &=& \omega_B b^\dagger b + \frac{U_{BB}}{2} b^\dagger b^\dagger 
b b,\\
H_{\rm Int} &=& U_1 a^\dagger b^\dagger b a,\\
H_{\rm Las} &=& -\lambda \left( a^\dagger b e^{-i\Delta t} + b^\dagger 
a
 e^{+i\Delta t}\right).
\end{eqnarray}
\end{mathletters}
This model corresponds to the previous one in the limit in which the
external motion of the atoms is frozen.  The 
bosonic annihilation and 
creation operators $a$,$a^\dagger$, $b$, $b^\dagger$ annihilate and 
create atoms in internal states $|A\rangle$, and $|B\rangle$, 
respectively. They fulfill standard commutation relations=
$[a,a^\dagger]=[b,b^\dagger]=1$. As before, we will assume
$\omega_B=\omega_A\equiv\omega$, $\Delta=0$,  and 
$U_{AA}=U_{BB}\equiv U_0$. This
allows us to neglect the first term in $H_A$ and $H_B$, since the total
number of particles $N=a^\dagger a + b^\dagger b$ is conserved. Note
 that the same simplification occurs when $\omega_A\ne\omega_B$, 
but
$\Delta=\omega_A-\omega_B$. This means that the results obtained in 
the next section
for $A-B$ symmetric case, are equivalent to  the ones
for the asymmetric case  with appropriately chosen $\Delta$.

An additional motivation behind the use of the model (\ref{Ham4}) is 
that it
can be solved numerically for moderate $N$, and therefore it allows to
compare the analytical approximations with the exact numerical results.
This will provide us with a guideline for
 the analysis of the complete quantum field theoretical
model section III.

\section{Analysis of the two--mode model}

In this section we consider in details the simple two--mode model
described by the Hamiltonians (\ref{Ham4}). The section is divided into
3 subsections. In the first of them we derive the ground state energy of
(\ref{Ham4})  using a mean field 
approach, for which all the atoms are
supposed to be in the same single particle state. In subsection III.B we
refine this theory to find a better approximation to the ground state.
We show that under certain conditions the ground state corresponds to a
{\em Schr\"odinger cat state}. Finally, in subsection III.C we
diagonalize the Hamiltonian exactly using a numerical method
(\ref{Ham4}), and compare the exact predictions with the approximate
ones of the previous subsections.

\subsection{Mean Field Approximation}

The equations for the ground state in the mean field (Hartree-like)
approximation can be derived using the standard procedure. We consider
the single particle state
\begin{equation}
\label{psi1}
|\psi_1\rangle = \alpha_1 |A\rangle + \beta_1 |B\rangle,
\end{equation}
where $|\alpha_1|^2+|\beta_1|^2=1$, and look for collective states
of $N$ particle system, with all the
particles in the same state (\ref{psi1}) which minimizes the total 
energy.
Using the  second quantization description, these collective states
can be represented as
\begin{equation}
\label{psiN}
|\psi_N\rangle = |\psi_1\rangle\otimes |\psi_1\rangle \ldots 
|\psi_1\rangle
= \frac{1}{\sqrt{N!}} \left[ \alpha_1 a^\dagger + \beta_1 b^\dagger
\right]^N |0\rangle,
\end{equation}
where $|0\rangle$ denotes the vacuum state.

The expectation value of the Hamiltonian
(\ref{Ham4}) in this state is
\begin{equation}
\label{mean}
E(\alpha,\alpha^\ast,\beta,\beta^\ast) = \langle \psi_N|H| \psi_N 
\rangle
= \frac{\tilde U_0}{2} \left(|\alpha|^4+|\beta|^4\right)
+ \tilde U_1 |\alpha|^2 |\beta|^2 - \lambda \left( \alpha^\ast
\beta + \beta^\ast \alpha \right),
\end{equation}
where we have redefined $\tilde U_{0,1}=
 U_{0,1} (N-1)/N$, and $\alpha=\sqrt{N}\alpha_1$ and 
$\beta=\sqrt{N}\beta_1$.
The
normalization condition imposes now
\begin{equation}
\label{normal1}
|\alpha|^2 + |\beta|^2 = N.
\end{equation}
For simplicity of the notation, we will drop the tilde over $U$'s
in the following.

We  minimize the mean energy   (\ref{mean}) with 
respect to
$\alpha,\beta$ and their complex conjugates,
imposing the constraint (\ref{normal1}) by
using a Lagrange multiplier $\mu$. After elementary calculations we 
obtain,
\begin{mathletters}
\label{lag}
\begin{eqnarray}
\left[ U_0 |\alpha|^2 + U_1 |\beta|^2 \right]\alpha - \lambda \beta 
&=& \mu \alpha,\\
\left[ U_0 |\beta|^2 + U_1 |\alpha|^2 \right]\beta - \lambda \alpha 
&=& \mu \beta.
\end{eqnarray}
\end{mathletters}
The above  equations can be easily solved. To this aim, we first note 
that
for $\lambda\ne 0$, $\alpha$ and $\beta$ can be taken to be 
 nonvanishing real numbers.
Thus, we can divide the first equation
by $\alpha$ and the second by $\beta$, and subtract them to obtain
\begin{equation}
\label{Dupa}
\left[U_1-U_0 - \frac{\lambda}{\alpha\beta} \right] (|\alpha|^2 
-|\beta|^2)=0.
\end{equation}

The analysis of Eq.~(\ref{Dupa}) is straightforward. Defining
$\Lambda=2\lambda/[N(U_1-U_0)]$ one finds that for $|\Lambda|>1$ there
exists only one solution
\begin{equation}
\alpha_0=\beta_0=\sqrt{N/2},
\end{equation}
which gives the  mean energy energy (\ref{mean})
\begin{equation}
E_0 = \frac{N^2}{4} (U_0 + U_1) -\lambda N;
\end{equation}
For $|\Lambda|<1$ there exist three solutions $(0,+,-)$:
\begin{mathletters}
\begin{eqnarray}
\alpha_0 &=& \beta_0=\sqrt{N/2}, \\
\label{alpm}
\alpha_\pm &=& \beta_\mp = \left[ \frac{N}{2}
\left(1\pm \sqrt{1-\Lambda^2}\right)\right]^{1/2},
\end{eqnarray}
\end{mathletters}
with the corresponding energies
\begin{mathletters}
\begin{eqnarray}
E_0 &=& \frac{N^2}{4} (U_0 + U_1) -\lambda N\\
E_+&=& E_- = U_0\frac{N^2}{2}-\frac{\lambda^2}{U_1-U_0}.
\end{eqnarray}
\end{mathletters}
One can easily check that for $U_0>U_1$ we have $E_\pm>E_0$
and therefore the solution $(\alpha_0,\beta_0)$ gives
the minimum energy. On the other hand, for $U_1>U_0$
both solutions $(\alpha_+,\beta_+)$ and $(\alpha_-,\beta_-)$ give
a lower energy than $(\alpha_0,\beta_0)$ (in particular, for 
$\Lambda=1$, $E_0=E_\pm$).

The results can be summarized as follows:

\begin{description}

\item {\it (a) Weak interactions between atoms in the states $|A\rangle$ 
and $|B\rangle$:}
In  this  situation  $U_0>U_1$ and the mean 
field wavefunction for the 
ground state is
\begin{equation}
\label{psi0}
|\psi_N^0\rangle = \frac{1}{\sqrt{2^N N!}} \left[ a^\dagger + 
b^\dagger \right]^N |0\rangle = 
\frac{1}{\sqrt{N^N N!}} \left[ \alpha_0 a^\dagger
+ \beta_0 b^\dagger \right]^N |0\rangle.
\end{equation}

\item {\it (b) Strong interactions between atoms in the states
$|A\rangle$ and  $|B\rangle$:}
In this situation $U_1>U_0$ and one has to 
distinguish two cases:

\begin{itemize}

\item {\it (b.1) Strong laser case:} $\Lambda\ge 1$ and the mean field 
wavefunction is $|\psi_N^0\rangle$
given in (\ref{psi0}).

\item {\it (b.2) Weak laser case:} $\Lambda < 1$; there are two 
degenerate solutions $|\psi_N^\pm\rangle$
for the mean field ground state wavefunction
\begin{equation}
|\psi_N^\pm\rangle = \frac{1}{\sqrt{N^NN!}} \left[ \alpha_\pm 
a^\dagger + \beta_\pm b^\dagger
\right]^N |0\rangle,
\end{equation}
where $\alpha_\pm$ and $\beta_\pm$ are given by expression  
(\ref{alpm}).

\end{itemize}
\end{description}

\subsection{Beyond the  Mean Field Approximation}

For the chosen  parameters, the Hamiltonian (\ref{Ham1}), with 
(\ref{Ham4}) is invariant under the
operation $T_{AB}$ that exchanges the internal level $|A\rangle$ with
$|B\rangle$. Thus, in case of no degeneracy the eigenstates of $H$,
$|\phi_k\rangle$ must be eigenstates of $T_{AB}$ too. Since $T_{AB}$ is
idempotent (i.e. has eigenvalues $\pm 1$), the eigenstates
have to fulfill
$T_{AB}|\psi_k\rangle = \pm |\psi_k\rangle$. For $|\alpha|\ne |\beta|$
(the case b.2 above), it is clear that the states obtained using the 
mean field
approach (\ref{psiN}) do not satisfy this condition. This indicates that
in the case (b.2) one can obtain a better approximation to the ground 
state
with a lower energy if one uses as a
variational ansatz the wavefunction
\begin{equation}
\label{psisch}
|\psi^\pm\rangle = \left(|\psi_N^+\rangle \pm 
|\psi_N^-\rangle\right)/\sqrt{2}.
\end{equation}
This is a superposition of the two degenerate solutions. Note that
$|\psi^\pm\rangle$ is indeed an eigenstate of $T_{AB}$ with eigenvalue 
$\pm 1$,
and therefore it conforms with the symmetry of the Hamiltonian.

The states (\ref{psisch}) are written as a superposition of two states 
in
which either all the atoms are in the single particle state 
$|\psi_1^+\rangle
= \alpha_1^+ |A\rangle + \beta_1^+ |B\rangle$, or all are in the 
single particle
state $|\psi_1^-\rangle = \alpha_1^- |A\rangle + \beta_1^- |B\rangle$.
Therefore, they have the form of Schr\"odinger cat states. 
Note, however, that a Schr\"odinger cat state is characterized by its coherent inclusion
of macroscopically distinguishable states. For the state of our
condensates to be a true (i.e. macroscopic, or at least mezoscopic)
Schr\"odinger cat state  we must therefore
require that  the
overlap $\epsilon$,
\begin{equation}
\epsilon = \langle \psi_N^+|\psi_N^-\rangle=\Lambda^{N},
\end{equation}
be as small as possible. The ``size of the cat'', which can be defined as $1/\epsilon$, should, on the other hand be as large as possible.
 The theory should determine under which conditions the
observation of the  ``cat of maximal size'' is feasible.

The expectation value of the energy of the state (\ref{psisch}) is
\begin{eqnarray}
E_{\pm} &=& \frac{\langle \psi_N^+|H|\psi_N^+\rangle \pm \langle
 \psi_N^+|H|\psi_N^-\rangle}
{1\pm \langle \psi_N^+|\psi_N^-\rangle}\nonumber\\
&=& \frac{N^2}{4}
\frac{2U_0-\Lambda^2(U_1-U_0)\pm (\Lambda)^{N}(3U_0-U_1)}{1\pm 
(\Lambda)^{N}}. \label{res20}
\end{eqnarray}
It is easy to check that in the limit of $\epsilon\simeq 1$ (i.e. when the cat is still microscopic), we obtain
\begin{equation}
\Delta E= \frac{N}{2}(U_1-U_0).
\end{equation}
This Eq. reveals characteristic scaling of the energy difference $\Delta E$
with $N$, which, as we shall see below, is also valid in the more interesting limit of  
$\epsilon=\Lambda^{N} \ll 1$ (i.e; when the cat is mezo-, or macroscopic). 
In such case we may first expand the result (\ref{res20})
in $\epsilon$, and then in 
$N\gg 1$, so that we obtain
\begin{equation}
\Delta E= E_- - E_+  -\simeq \epsilon \ln (\epsilon) N (U_1-U_0).
\end{equation}
Thus, for a ``given size'' of the cat $1/\epsilon$ the energy difference
is proportional to $N$. Quite generally, the difficulty in cooling to a
ground state of a {\em given} purity increases with growing number of
atoms $N$, while a larger energy gap $\Delta E$ makes the cooling
easier. In this sense, the scaling $\Delta E \propto N$ helps. 
%

\subsection{Exact Numerical Solution}

In the Fock basis $|m\rangle_A\otimes |N-m\rangle_B$ 
($m=0,1,\ldots,N$)
the Hamiltonian $H$ is a $(N+1)\times(N+1)$ tridiagonal matrix, and can
therefore be easily diagonalized by numerical methods. Since the mean
field approximation and its improved version analyzed in previous
subsections should be valid in the limit $N\rightarrow \infty$, we
concentrate here on the finite $N$ results.

Let us  denote by
\begin{equation}
\label{dupupa}
|\phi_k\rangle = \sum_{m=0}^N q^k_m |m\rangle_A\otimes 
|N-m\rangle_B,
\end{equation}
the eigenstate corresponding to the energy $E_k$ ($k=0,1,\ldots,N$,
and $E_0\le E_1\le \ldots<E_N$). The results of our analysis are 
presented in
Figs. 1-4. In Fig.~1  we have plotted the  ground state energy $E_0$ as 
a function of
$\Lambda$ for $N=1000$ and 
$U_1=3U_0$. Although this Figure already shows the clear signatures of 
the ``phase transition'' to the ``Schr\"odinger cat''phase for 
$\Lambda<1$, it is more instructive to look at
relative behavior of the consecutive eigenenergies  of the low excited 
states.
This is represented in Fig. 2(a) for $N=1000$ and 2(b) for 
$N=10000$, where
we have plotted the ratio between the energy difference of
the first excited state and the ground state, and
 the energy difference of the second and first excited states
 $(E_1-E_0)/(E_2-E_1)$,
as a function of $\Lambda$ for $U_1=3U_0$. The inset in Fig. 2(b) 
shows the magnification of the transition region. The figures clearly 
show that as
$\Lambda$ becomes smaller than 1, the energies of the first excited and 
the ground state merge together. These two states become 
quasi-degenerate, whereas the energy gap to the second excited state 
remains finite. Since the ground state is even, and the first excited 
state is odd with respect to the $A-B$ symmetry,
and since they both are Schr\"odinger cat states,
their sum or difference describe the ``dead'' or ``alive cat'', 
respectively.

In Fig. 3(a) and 3(b)  we have plotted the energy of the first, 
second and third
excited states with respect to the energy of the ground state as
a function of $\Lambda$ for $N=1000$ and $N=10000$, respectively.
This figures clearly illustrate that, as expected, merging of the energy levels 
occurs not only for the two lowest ones, but also  within
consecutive pairs of levels, i.e. $E_3$ becomes practically equal to 
$E_2$ for $\Lambda<1$, etc.

Finally, in  Fig.~ 4  we have plotted the $A$-atom number distributions 
for the ground state (i.e. the coefficients $|q_m^0|^2$ from Eq. 
(\ref{dupupa}) as a function of $m$)  for $N=1000$ [Fig.~4(a)]
and $N=10000$ [Fig.~4(b)] for different values of $\Lambda$. These 
values belong to the transition regions  between the dashed lines  in 
Figs.~3(a) and 3(b).

The comparison of these results with the mean field theory and its
improved version is very satisfactory. Mean field theory is practically
exact outside of the transition region, and gives errors of the order of
$O(1/N)$. The improved mean field approximation of the subsection III.B
does the similar job for all values of the parameters, i.e. including
the transition region. This result indicates that a similar improved
mean field approach can be used for the complete field theoretical
model. We adopt this approach in the next section.

Finally, the results indicate that due the finite energy gap between the
ground and first excited state it is possible to prepare and detect the
Schr\"odinger cat state in the following manner. Obviously, direct
cooling of the system to the absolute ground state, which for $\Lambda <
1$ is a cat state, would be a hardly possible task. The idea is
therefore to first cool the system to a temperature $T$ close to zero
(i.e. such that $(E_1-E_0)/k_B T<1$) for $\Lambda>1$. Note, that this is
only possible in this regime of $\Lambda$, since only there the first
excited state energy is high enough, so that practically all of the
atoms can be cooled down to the ground state. Then we decrease $\Lambda$
($\lambda$) adiabatically and enter the ``Schr\"odinger cat phase'': the
system remains in the ground state, which now becomes the
``Schr\"odinger cat'' state. 

Internal state selective
atom counting would then reveal a two-peaked structure corresponding to 
the
$|\psi_N^{\pm}\rangle$ component. This, of course, would not prove the 
coherence.
In the most general case, this would require tomographic techniques
to reconstruct the complete density matrix of the two mode system, 
similar to
those developed for photon fields \cite{Walser}.

\section{Analysis of the quantum field theoretical model}

Here we analyze the full problem described by the Hamiltonians
(\ref{Ham1}) and (\ref{Ham3}) that account for the atomic motional
degrees of freedom. Given the similarities of this model to the two-mode
model analyzed above, we follow  similar steps as in section III. In
the first subsection, we apply the mean field approximation to
characterize the ground state of the full Hamiltonian. In principle, the
exact solution of the mean field equations is already very difficult to
handle and requires numerical treatment. We have used instead two
different methods to analyze it: the Thomas--Fermi approximation in
subsection IV.C, and the Gaussian variational ansatz for the single
particle wavefunction in subsection IV.D (for both methods c.f.
\cite{baym}). In both cases we find qualitatively the same results
as for the two-mode model; in particular, in the strong 
interaction and low intensity limit (the case b.2 above) there are two
degenerate solutions of the mean field equations. In the subsection IV.E
we go beyond the mean field approximation to analyze these results.
Finally, in the last subsection we utilize an even more sophisticated
model to approximate numerically the eigenstates of the system.

\subsection{Mean Field Approximation}

As in subsection III.A,  we assume that the ground state of the system 
is a state
  for which all the atoms are in the same single particle
state
\begin{equation}
\langle \vec r|\psi_1\rangle= \alpha_1(\vec r) |A\rangle + 
\beta_1(\vec r) |B\rangle,
\end{equation}
where
\begin{equation}
\label{Normal2}
\int d^3\vec r \left( |\alpha_1(\vec r)|^2 + |\beta_1(\vec r)|^2 \right) 
= 1.
\end{equation}
The collective ground state of the whole system will then be, using the 
second quantization
description,
\begin{equation}
\label{Total}
|\psi_n\rangle = |\psi_1\rangle\otimes |\psi_1\rangle \ldots 
|\psi_1\rangle
= \frac{1}{\sqrt{N!}} \left[ \int d^3\vec r \left[\alpha_1(\vec 
r)^\ast
\hat\Psi_A(\vec r)^\dagger
+ \beta_1(\vec r)^\ast \hat\Psi_B(\vec r)^\dagger \right]\right]^N 
|0\rangle,
\end{equation}
where $|0\rangle$ denotes the vacuum state.
The mean energy
of  this state can be easily calculated,
\begin{eqnarray}
\label{mean2}
E(\alpha,\alpha^\ast,\beta,\beta^\ast)& = &\langle \psi_N|H| \psi_N 
\rangle
= \frac{1}{2} \int d^3\vec r \alpha(\vec r)^\ast
\left[  -\frac{\nabla^2}{2} + \frac{r^2}{2}
+ \frac{ U_0}{2} |\alpha(\vec r)|^2 + \frac{ U_1}{2} |\beta(\vec 
r)|^2\right]
\alpha(\vec r) \nonumber \\
&-& \frac{\lambda}{2} \int d^3\vec r \left[\alpha(\vec r)^\ast 
\beta(\vec r)
\beta(\vec r)^\ast \alpha(\vec r)\right] + (\alpha 
\leftrightarrow\beta).
\end{eqnarray}
Here, as in the case of the two--mode system, we have defined $ \tilde
U_0= U_0(N-1)/N$, $\tilde U_1= U_1(N-1)/N$ (for simplicity we will 
omit
the tilde in the following),  and $\alpha(\vec
r)=\sqrt{N} \alpha_1(\vec r)$ and $\beta(\vec r)=\sqrt{N} 
\beta_1(\vec
r)$. The normalization condition (\ref{Normal2}) becomes now
\begin{equation}
\label{Normal3}
\int d^3\vec r \left( |\alpha(\vec r)|^2 + |\beta(\vec r)|^2 \right) = 
N.
\end{equation}

In Eq. (\ref{mean2}) the expectation value of the energy is expressed as 
a
functional of the single particle wavefunctions $\alpha(\vec r)$ and
$\beta(\vec r)$. The goal is now to minimize this energy with respect to
these functions. In general, it is a difficult task that can be treated
only numerically. In the next
subsections we will follow two different approaches to find the
solutions to this problem: first, we will analyze the Thomas--Fermi
limit, and second we will use Gaussian ansatz.

\subsection{Thomas--Fermi approximation}

In order to minimize (\ref{mean2}) we calculate  the functional 
derivatives of
the mean energy $E(\alpha,\alpha^\ast,\beta,\beta^\ast)$ with respect to
$\alpha,\beta$, and their complex conjugates using a Lagrange multiplier
$\mu$ to ensure that the normalization condition (\ref{Normal3}) is 
fulfilled.
This leads to a set of coupled nonlinear Sch\"odinger equations,
\begin{mathletters}
\label{dupa2}
\begin{eqnarray}
\left[- \frac{\nabla^2}{2} + \frac{r^2}{2} +  U_0 |\alpha(\vec r)|^2
+  U_1 |\beta(\vec r)|^2 \right] \alpha(\vec r) - \lambda \beta(\vec r) 
&=&
\mu \alpha(\vec r),\\
\left[- \frac{\nabla^2}{2} + \frac{r^2}{2} +  U_0 |\beta(\vec r)|^2
+  U_1 |\alpha(\vec r)|^2 \right] \beta(\vec r) - \lambda \alpha(\vec r) 
&=&
\mu \beta(\vec r).
\end{eqnarray}
\end{mathletters}
These equations are equivalent to Eqs. (\ref{lag}) for the two--level 
model.
In the Thomas--Fermi approximation one assumes that the  
terms involving  $\nabla^2$ can be neglected in comparison with the
interaction and potential terms.

According to Eqs. (\ref{dupa2}), for $\lambda\ne 0$, at any position
$\vec r$, if $\beta(\vec r)=0$ then $\alpha(\vec r)=0$.
This can be understood as follows: consider, for example, that at
some point $\alpha(\vec r)\ne 0$ and  $\beta(\vec r)= 0$, then the 
laser will
take particles from the state $|A\rangle$ to the state $|B\rangle$, 
which will imply
that $\beta(\vec r) \ne 0$.
This is not the case if $\lambda=0$, where displaced solutions in the
Thomas--Fermi limit are indeed possible \cite{tfbi}. Thus, we can
 concentrate on the positions $\vec r$
where $\alpha(\vec r),\beta(\vec r)\ne 0$. Dividing (\ref{dupa2}) by 
 $\alpha(\vec r)$
and
$\beta(\vec r)$ respectively, and taking the difference we find
\begin{equation}
\label{other}
\left[  U_0 -  U_1 + \frac{\lambda}{\alpha(\vec r)\beta(\vec r)} \right]
\left[ |\alpha(\vec r)|^2-\beta(\vec r)|^2 \right] =0,
\end{equation}
which resembles very much Eq. (\ref{Dupa}). The analysis is, however, a 
bit
more complicated. As before, there are two kinds of solutions, 
$|\alpha(\vec r)|=|\beta(\vec r)|$
and $|\alpha(\vec r)|\ne |\beta(\vec r)|$, where the  latter exists
for sufficiently small $\lambda$ only. In more detail, we can 
distinguish
the following cases:

\begin{description}

\item {\it (a) Weak interactions between atoms in the states $|A\rangle$ 
and $|B\rangle$:}
In this situation  $U_0>U_1$ and the mean field wavefunction for the 
ground state is
(\ref{Total}) with
\begin{equation}
\label{sol1}
\alpha(\vec r) = \beta(\vec r) =\sqrt{\frac{1}{2( U_0+ U_1)}
\left(r_0^2-r^2\right)},
\end{equation}
where (for an isotropic trap in 3D)
\begin{equation}
r_0=\left[ \frac{15}{8\pi} N ( U_0+ U_1)\right]^{1/5}.
\end{equation}

\item {\it (b) Strong interactions between atoms in the states  
$|A\rangle$ and $|B\rangle$:}
In this situation $U_1>U_0$, and one has to distinguish two cases:

\begin{itemize}

\item {\it (b.1) Strong laser case:} For
\begin{equation}
\Lambda\equiv \frac{2\lambda}{U_1-U_0} \ge \Lambda_0\equiv
\left[\frac{15N}{8\pi}\right]^{2/5} (U_1+U_0)^{-3/5}
\end{equation}
the mean field ground state wavefunction is the same as in the case (a).

\item {\it (b.2) Weak laser case:} For $\Lambda < \Lambda_0$ there exist
two degenerate solutions $|\psi_N^\pm\rangle$ for the mean field ground
state wavefunction of the form (\ref{Total}) with the coefficients
$\alpha(\vec r)$, and $\beta(\vec r)$ given by
\begin{mathletters}
\label{sol2}
\begin{equation}
 \alpha_\pm(\vec r)^2 = 
\beta_\mp(\vec r)^2 = \frac{1}{\sqrt 2} \left[\rho_\mu(\vec r) \pm
\sqrt{ \rho_\mu(\vec r)^2- \Lambda^2} \right]^{1/2},
\end{equation}
 for $r\le r_1\equiv\sqrt{2(\mu-\Lambda U_0)}$, and
\begin{equation}
\alpha_+(\vec r)=\alpha_-(\vec r)=\frac{U_1-U_0}{U_1+U_0}
\left[\rho_\mu(\vec r) +\Lambda/2\right],
\end{equation}
for  $r_2\equiv\sqrt{2\mu+(U_1-U_0)\Lambda)}\ge r\ge r_1$,
\end{mathletters}
Here, $\rho_\mu(\vec r)\equiv(\mu-r^2/2)/ U_0$, and the
value of the Lagrange multiplier $\mu$ has to be found by imposing
the constraint (\ref{Normal3}).
\end{itemize}
\end{description}

Apart from numerical factors and different scalings, the Thomas--Fermi
approximation gives results which are qualitatively similar to those
found for the
simple two--level model.

\subsection{Gaussian Variational ansatz}

The Thomas--Fermi solution found in the previous Section is valid for
$N\rightarrow\infty$ (or strictly speaking $NU_{0,1}\rightarrow\infty$),
and predicts the existence of degenerate solutions for sufficiently low
laser intensities. It is interesting to see if the   effects remain
for finite $N$. This can be analyzed using a Gaussian variational
ansatz for the wavefunctions, i.e. setting
\begin{mathletters}
\begin{eqnarray}
\alpha(\vec r) &=& \sqrt{A} e^{-r^2/(4a)},\\
\beta(\vec r) &=& \sqrt{B} e^{-r^2/(4b)},
\end{eqnarray}
\end{mathletters}
with the variational parameters $A$, $B$, $a$, $b$.
These parameters are not completely independent, since the normalization 
(\ref{Normal3}) requires
\begin{equation}
\label{Norm4}
N=(2\pi)^{3/2} (Aa^3+Bb^3).
\end{equation}
Substituting this ansatz in Eq. (\ref{mean2}), we find that the
expectation value of the energy depends on the variational
parameters $a,A,b,B$. We  minimize it then with respect to
those parameters taking into account the normalization condition
(\ref{Norm4}).

We have not found analytical solutions in this case. However, we have
solved the problem numerically, and found the same qualitative results
as in the Thomas--Fermi approximation. That is, in the case of weak
interactions between atoms in the states $|A\rangle$ and $|B\rangle$
($U_0>U_1$) there exists only one solution which corresponds to $A=B$ 
and
$a=b$. In the case of strong interactions between atoms in the states 
$|A\rangle$
and  $|B\rangle$ ($U_1>U_0$), for a given number
of particles $N$ we find that there exists  $\Lambda_0(N)$ such that if
$\Lambda > \Lambda_0$
the minimum energy corresponds to the case $A=B$ and $a=b$ again;
Conversely, if $\Lambda\le \Lambda_0$ there exist two
solutions with $A\ne B$ and $a\ne b$.

\subsection{Beyond the Mean Field}

For our choice of parameters, the Hamiltonian (\ref{Ham1}), with
(\ref{Ham3}) is invariant under the operation $T_{AB}$ that exchanges
the internal level $|A\rangle$ with $|B\rangle$. Thus, the ground state
of the Hamiltonian has to be an eigenstate of this operator. If we
impose this condition  on the ansatz (\ref{Total}) we find that
$|\alpha(\vec r)|=|\beta(\vec r)|$. As we have seen in the previous
subsections (case b.2), for a given number of particles there exists a
certain $\Lambda_0$ such that if $\Lambda<\Lambda_0$ there are two
wavefunctions of the form (\ref{Total}) $|\psi_\pm\rangle$ with
$\alpha_\pm(\vec r)\ne \beta_\pm(\vec r)$ that have lower energy than
that given by the solution $|\alpha(\vec r)|=|\beta(\vec r)|$. This
implies in turn that there exists a better variational ansatz to the
problem, namely
\begin{equation}
|\psi^\pm\rangle = |\psi_N^+\rangle \pm |\psi_N^-\rangle,
\end{equation}
where
\begin{equation}
|\psi_N^\pm\rangle
= \frac{1}{\sqrt{N!}} \left[ \int d^3\vec r \left[\alpha_1^\pm(\vec 
r)^\ast
\hat\Psi_A(\vec r)^\dagger
+ \beta_1^\pm(\vec r)^\ast \hat\Psi_B(\vec r)^\dagger\right] \right]^N ,
\end{equation}

The corresponding energy is given by 
\begin{equation}
E_{\pm} =\frac{\langle \psi_N^-|H|\psi_N^+\rangle
 \pm \langle \psi_N^+|H|\psi_N^-\rangle}
{1\pm \langle \psi_N^+|\psi_N^-\rangle}.
\end{equation}
It can be easily checked that $E_+< \langle \psi_N^+|H|\psi_N^+\rangle < 
E_-$.
Thus, similarly  to the two--level model, the proper
ground state ansatz has the form of
a Schr\"odinger cat state.

\subsection{Approximate Numerical Solution}

It is possible to use an even more general ansatz to generate better 
approximations
to the real ground state of the Hamiltonian (\ref{Ham1}). In that
case no analytical approximation is possible, and one has to restrict
oneself to numerical evaluations. In any case, one can check whether
the existence of Schr\"odinger cat   states is compatible with these
more elaborated calculations,  and on may confirm the mean field
solutions. In the following we use the ansatz 
\begin{equation}
\label{Psi}
|\hat\Psi\rangle = \sum_{m=0}^N \frac{q_m}{\sqrt{m!(N-m)!}}
\left[ \int d^3\vec r \alpha_m(\vec r) \hat\Psi_a(\vec 
r)^\dagger\right]^{m}
\left[ \int d^3\vec r \beta_{N-m}(\vec r) \hat\Psi_b(\vec 
r)^\dagger\right]^{N-m} |0\rangle,
\end{equation}
where $q_m$'s and the wave functions $\alpha_m(\vec r)$, and
$\beta_m(\vec r)$ are the variational parameters. To conform to the
symmetry of the full Hamiltonian we impose additionally that
\begin{equation}
\beta_m(\vec r)= \alpha_m(\vec r),\quad\quad q_m = q_{N-m}.
\end{equation}

If the ansatz (\ref{Psi}) is used to calculate the expectation
value of the Hamiltonian $\langle \hat\Psi|H|\hat\Psi\rangle$, one finds
a rather complicated expression
 involving the expansion coefficients $q_m$ and the 
wavefunctions $\alpha_m(\vec r)$.
an infinite set of nonlinear Schr\"odinger equations 
that couples $q_m\rightarrow q_m,q_{m\pm 1}$
as well as  $\alpha_m \rightarrow \alpha_m,\alpha_{m\pm 1},
\alpha_{N-m},\alpha_{N-m\pm 1}$.
A solution of these  equations seems to be an impossible task,
but fortunately the  equations simplify in the limit of  
sufficiently large $N$. We have proved using the systematic $1/\sqrt{N}$ 
expansion, that
in this limit one can simply substitute $m\simeq m\pm 1$,
which implies  $\alpha_m\simeq \alpha_{m\pm 1}$, as well as $q_m\simeq 
q_{m\pm 1}$ in the equations for $\alpha_m$'s.

The resulting set of differential equations for $\alpha_m(\vec r)$'s
has the following form:
\begin{equation}
\label{vari}
\left[-\frac{\nabla^2}{2}+\frac{r^2}{2}+U_0 |\tilde\alpha_m(\vec r)|^2
+U_1 |\tilde \alpha_{N-m}(\vec r)|^2\right] \tilde\alpha_m(\vec r)
-2 \lambda\tilde \alpha_{N-m}(\vec r) = \mu_m \tilde\alpha_m(\vec r)
\end{equation}
where $\tilde \alpha_m(\vec r)\equiv \sqrt{m}\alpha_m(\vec r)$, and
with $\mu_m$ such that the normalization condition
\begin{equation}
m=\int d^3\vec r |\tilde \alpha_m(\vec r)|^2
\end{equation}
is fulfilled.
The above equations have to be accompanied by the linear Schr\"odinger 
equations for $q_m$ that contain tri-diagonal coupling to $q_{m\pm 1}$. 
The coefficients in the latter  equations depend, however, functionally 
in a highly nonlinear way on  the $\alpha_m$'s,
\begin{equation}
Eq_m=E_mq_m -\lambda L_mq_{m-1}-\lambda K_mq_{m+1},
\end{equation}
where $E$ denotes  the eigenvalue we search for, whereas
\begin{eqnarray}
E_m&=&  \frac{1}{2} \int d^3\vec r \alpha_m(\vec r)^\ast
\left[  -\frac{\nabla^2}{2} + \frac{r^2}{2}
+ \frac{ U_0}{2} |\alpha_m(\vec r)|^2 + \frac{ U_1}{2} 
|\alpha_{N-m}(\vec r)|^2\right]
\alpha_m(\vec r)  + (m \leftrightarrow N-m), \\
L_m&=&\sqrt{m(N-m+1)} \int d^3\vec r
\alpha_{m-1}(\vec r)^\ast \alpha_{N-m+1}(\vec r), \\
K_m&=&\sqrt{(m+1)(N-m)} \int d^3\vec r \alpha_{m+1}(\vec r)^\ast 
\alpha_{N-m-1}(\vec r).
\end{eqnarray}
The coefficients $q_m$ are normalized as
\begin{equation}
\sum_{m=0}^N |q_m|^2=1.
\end{equation}

Note that Eqs. (\ref{vari}) are similar to Eqs. (\ref{dupa2}) except that they
take fully into account the change of the form of the wave function with 
$m$.
Unfortunately, these equations are still very difficult to solve, even
using the Thomas--Fermi approximation.
We can show, however, that 
the condition $\alpha_{N-m}(\vec r)
> \alpha_m(\vec r)$ ($m<N/2$) is fulfilled everywhere, a
conclusion that could also be reached in the context of 
our mean field theory.
Guided by this observation, we have used simple Gaussian
functions  to approximate the
solutions of Eqs. (\ref{vari}).  In other words, we have
set
\begin{equation}
\label{gaucho}
\alpha_m(\vec r) = \sqrt{A_m} e^{-r^2/(4a_m)},
\end{equation}
and minimize the mean field energy numerically with respect to
the $a_m$. Note, that the normalization condition implies
automatically that $m=A_m(2\pi a_m^2)^{3/2}$, so that the value of 
$A_m$ is
determined by the value of $a_m$. In an even more sophisticated attempt
we have used as a  variational ansatz a sum of two different Gaussians
of the form  (\ref{gaucho}). This calculation has led to
practically the same results as the ones obtained with a single Gaussian 
ansatz.
For this reason we present below numerical results corresponding to a 
single Gaussian ansatz.

Our main results are shown in Figs. 5 and 6. Fig. 5 is the
straightforward analog of Fig. 2(a). We have plotted there the ratio
between the energy difference of the first excited state and the ground
state, and the energy difference of the second and first excited states
$(E_1-E_0)/(E_2-E_1)$, as a function of $\Lambda$ for $N=1000$ and
$U_1=3U_0$. We have used the  parameters: $a^{\rm
sc}_A=a^{\rm sc}_B=50$ nm, $a^{\rm sc}_{AB}=150$ nm, $x_0=3$ 
$\mu$m,
$\hbar\omega=100$ Hz. The figure clearly shows that as in the the case
of the two-mode model as $\Lambda$ becomes smaller than 1, the energies
of the first excited and the ground state merge together. These two
states become quasi-degenerate, whereas the energy gap to the second
excited state remains finite.

Similarly, Fig. 6 is an analog of Fig. 3(a). There we have plotted the
energy of the first, second and third excited states with respect to the
energy of the ground state as a function of $\Lambda$. As already seen
in the  two-mode model, merging of the energy levels occurs also among consecutive 
pairs of levels, i.e. $E_3$ becomes practically equal to
$E_2$ for $\Lambda<1$.

In conclusion we stress that the similarity between the Figs. 2(a), 3(a)
and 5, 6 is remarkable. Clearly, the complete field theoretical model
leads to the same physics as the two-mode model. As $\Lambda$ is
adiabatically decreased, the system enters the ``Schr\"odinger cat
phase'' in which the ground state is a linear superposition of two
states for which the $A$-atom number distributions are significantly
distinct.

\section{Is a Schr\"odinger cat state experimentally feasible?}

We conclude with a summary of requirements to observe Schr\"odinger cats
in an experiment. While these necessary conditions to prepare and 
preserve
cat-like states are not fulfilled in
the present generation of Bose-Einstein experiments \cite{bicon}, they
might provide a guideline and motivation for future experimental work.

(i) {\em $A-B$ symmetry.} The calculations in the present paper assume
an $A-B$ symmetry. We believe that this assumption is mainly a technical 
point
in the theoretical calculation, but
discuss this now in more detail.

The choice of equal scattering length, $a^{\rm sc}_A=a^{\rm sc}_B$, is
reasonable, and agrees with the recent theoretical calculations
\cite{julliene}. The assumption of equal trap frequencies,
$\omega_A=\omega_B$, however, is typically not fulfilled in magnetic
traps, since atoms in different internal states feel different
(magnetic)
potentials, but could be realized in principle in an optical dipole
trap,  where it is assumed that the two states have the same
electron configuration, so that the far off resonant lasers induce the
same lightshifts. 
Another way of achieving equal trap frequencies, is to 
compensate an
asymmetry $\omega_A \ne \omega_B$ using an appropriately detuned laser.
Such compensation is exact in the case of the two-mode model. In the
case of the complete field theoretical model it requires a little more
care. The idea is that the necessary and sufficient condition for a
ground state of the system to be a Schr\"odinger cat state, is that
there exist two distinct and degenerate minima of the energy in the mean
field approximation. If $\omega_A\ne\omega_B$ the mean field theory
would typically lead to two minima of the energy function, but with
slightly different energies. The reader can easily convince him/herself
that this is the case for the two-mode model, while the analysis for the
complete model is more technical, but otherwise analogous (especially in
the Thomas-Fermi limit). That means that in such  cases we would have one
global, and one local minimum of the energy, and neither of them would
correspond to a ``Schr\"odinger cat'' state. They will, however, be
characterized by different numbers of atoms in the states $|A\rangle$
and $|B\rangle$. The point is  that the energies of these 
minima, and the
corresponding $A$- and $B$-atom numbers can be deformed in a continuous
manner by changing $\Delta$. At some point one arrives at the situation
when the two minima become degenerate, and at which the true ground
state becomes a linear combination of the two, i.e. becomes a
``Schr\"odinger cat''.

(ii) {\em Instability condition $a^{\rm sc}_{AB}>a_{A,B}^{\rm sc}$.}
The existence of cat states
imposes  the (instability) condition $a^{\rm sc}_{AB}>a_{A,B}^{\rm sc}$
(i.e. $U_1>U_0$). As we know, the present experiments \cite{bicon} with
Rb atoms allow for simultaneous evaporative and  sympathetic cooling
because the inelastic collision rates are small, which is directly
related to the fact that $a^{\rm sc}_{AB}\simeq a_{A}^{\rm sc}\simeq
a_{B}^{\rm sc}$ \cite{julliene}. While the condition  seems not to be 
satisfied
for atom species used in present {\em magnetic trap} experiments,
 we stress that future experiments might be based on different traps,
for example, far--off resonance traps using hightly detuned laser light.
This will open up the possibility of trapping new internal atomic states.
Consider for example a total angular momentum $F=1$, and condensates
in the states $m_F=\pm 1$. Furthermore, we assume that the level $m_F=0$
does not participate in the collision dynamics (this can be done, for
example, by shifting it with a laser). If the singlet scattering length
is larger than the triplet scattering length then the condition 
$a^{\rm sc}_{AB}>a_{A,B}^{\rm sc}$ will be fulfilled \cite{Stoof}.
In addition, in principle, to modify the atom-atom scattering 
lengths
\cite{kagan}.

(iii) {\em Cooling to the ground state.} Preparation of a ``cat''
requires the cooling to the ground state of our system, i.e the 
preparation
of a {\em pure} state.
The sufficient condition is that the temperature
has to be $\kappa_B T \alt E_1$, where $E_1$ is the energy of the first
excited state of the total Hamiltonian (for example, in the ideal case,
that would require that more or less $N-1$ particles are in the ground
state, and one is in the first excited state). We stress that this
requirement is much stronger than the requirement of having most of the
atoms in the single particle ground state, i.e obtaining a {\em
macroscopic occupation} of the ground state, as observed in current BEC
 experiments \cite{BEC-Rb,Ketterle,BEC-Li}.

We illustrate this by an example: if we have $N$ particles and 80\% of 
them are
in the ground state, the population of the collective ground state is
only $0.8^N$. Thus, the temperatures required to observe a ``cat'' are
much lower, such that practically all atoms are in the collective ground
state. Wether existing cooling techniques, in particular evaporative
cooling, can be extended into this regime remains to be investigated.
However, the fact that we are dealing with bosons instead of
distinguishible particles helps significantly to prepare
 a pure ``cat'' ground state.
Consider a one dimensional situation. For $N$ distinguishible
particles, there are $N$ possible collective excited states with the
same energy. Thus, the temperature required to have most of the
population in the collective ground state is $\kappa_B T \alt E_1/N$. On
the contrary, for bosons this temperature is {\em $N$ times higher},  
$\kappa T \alt E_1$.

(iv)  {\em Decoherence}. Finally, we should address the question of
decoherence. Obviously, atom losses (such as those due to inelastic
collisions) would destroy the ``Schr\"odinger cat'' state very rapidly.
In fact, in the extremal case when one has the cat
$|0,N\rangle+|N,0\rangle$, already one atom loss would be enough to
distort completely the coherent superposition (c.f.
\cite{zurek,modmea,brune}). We stress, however, that the situation here
is similar to that of the experiment of Brune {\it et al.} \cite{brune}.
The ``cats'' that live long enough to be observed must be mesoscopic. In
fact, the ``Schr\"odinger cat'' states displayed in Fig. 4 belong to
that category. They allow for loss of many atoms without the
complete smearing out of their quantum mechanical coherence. If they are
created, they could allow for the study of the gradual decoherence
process, in a similar manner as has been done in Ref. \cite{brune}.

\section{Acknowledgments}

We thank Y. Castin and H. Stoof for discussions. 
M.L. and K. M. thank J. I. Cirac and P. Zoller for hospitality extended
to them during their visit at the University of Innsbruck. 
Work at the
Institute for Theoretical Physics, University of Innsbruck was supported
by TMR network ERBFMRX-CT96-0002 and the \"Osterreichische Fonds zur
F\"orderung der wissenschaftlichen Forschung.

\begin{figure}
\centerline{
\epsfig{file=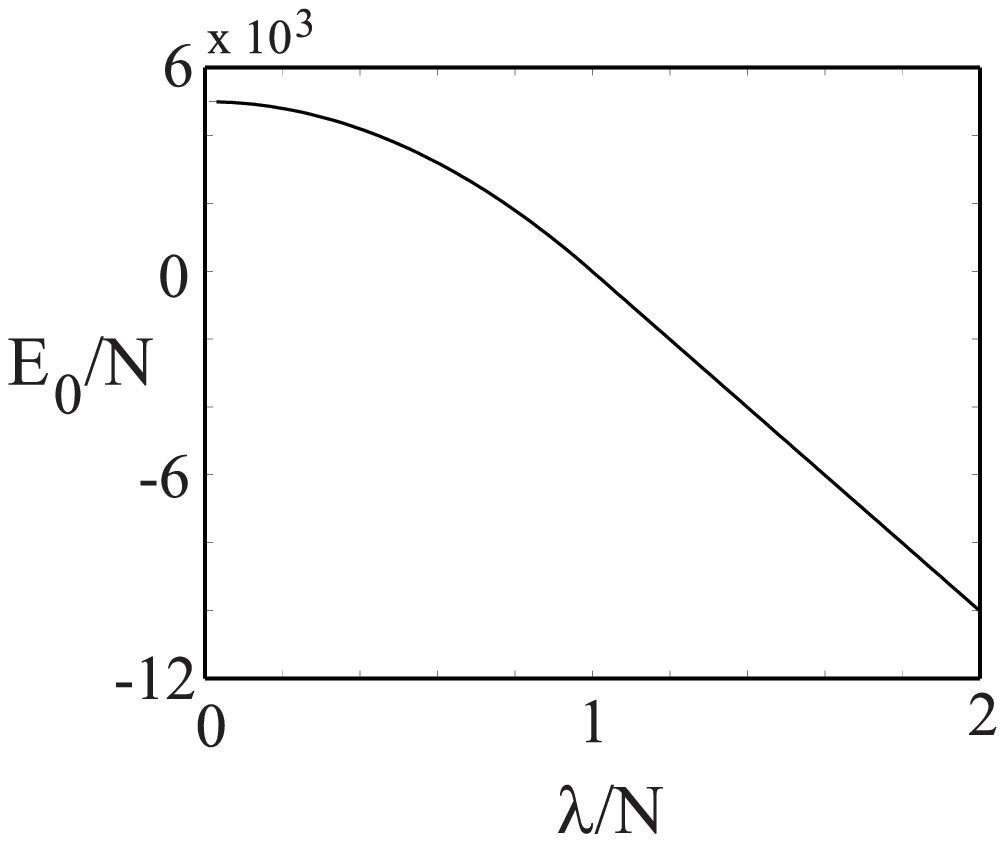,height=6.3 cm}}
\caption{The ground state energy $E_0$ as a function of
$\Lambda$ for $N=1000$ and
$U_1=3U_0$ calculated for the two--mode model. } 
\end{figure}

\begin{figure} 
\centerline{
\epsfig{file=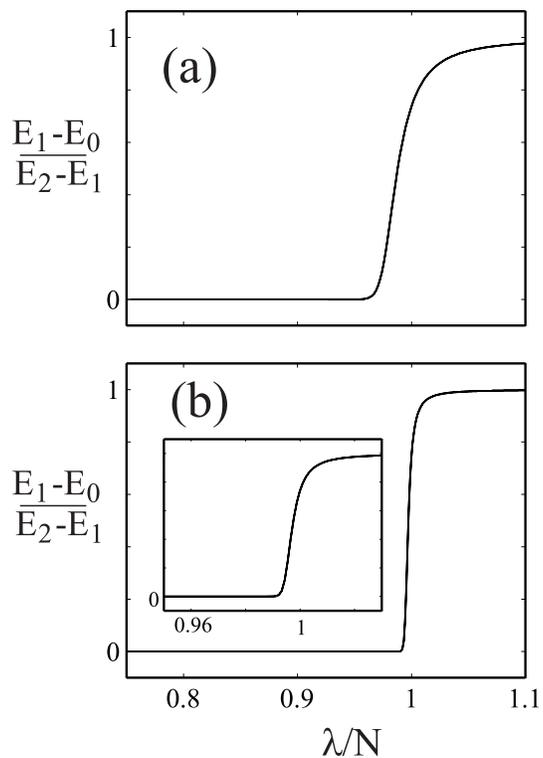,height=10 cm}}
\caption{ (a) The ratio between the energy difference of
the first excited state and the ground state, and the energy difference
of the second and first excited states $(E_1-E_0)/(E_2-E_1)$, as a
function of $\Lambda$ for $U_1=3U_0$ and for $N=1000$; (b) Same as 
(a),
but for $N=10000$. The inset shows the magnification of the transition
region.} 
\end{figure}

\begin{figure}\vspace*{0.8cm}
\centerline{
\epsfig{file=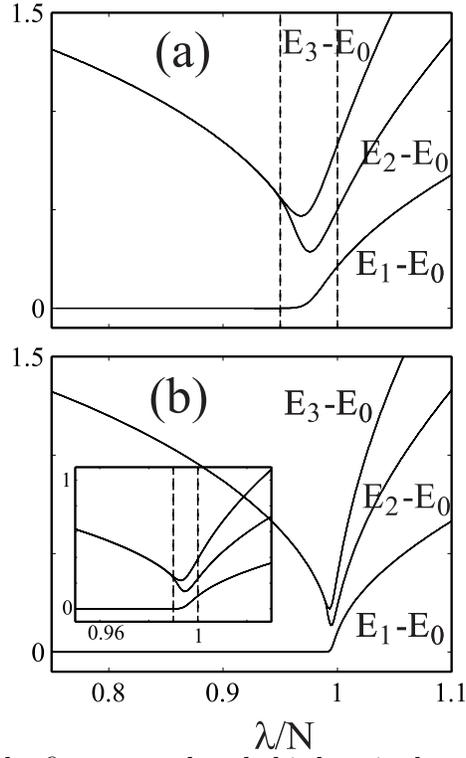,height=10 cm}}
\caption{(a) The energy of the first, second and third
excited states with respect to the energy of the ground state as a
function of $\Lambda$ for $N=1000$, and other parameters the same as 
in
Fig. 1. The dashed lines mark approximately the transition region; (b)
Same as (a), but for $N=10000$. The inset shows the magnification of 
the
transition region. }
 \end{figure}

\begin{figure} 
\centerline{
\epsfig{file=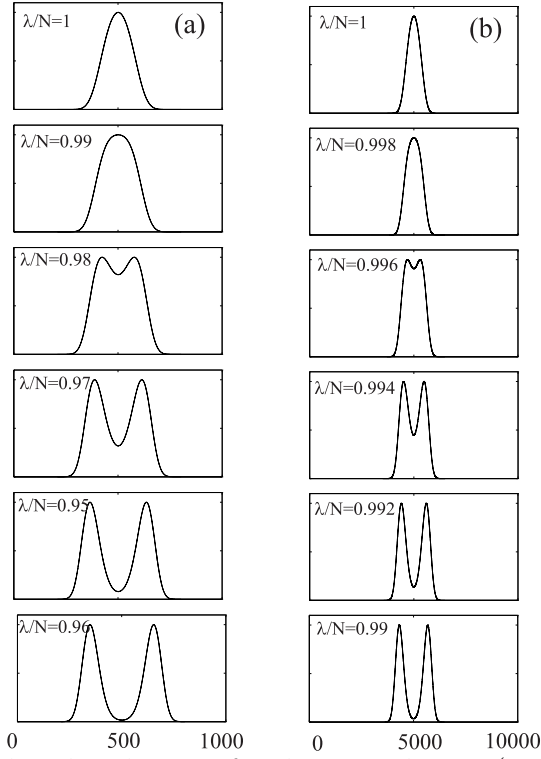,height=10 cm}}
\caption{(a) $A$-atom number distributions for the ground
state (i.e. the coefficients $|q_m^0|^2$ from Eq. (\ref{dupupa}) as a
function of $m$) for $N=1000$ for the values of $\Lambda$ indicated.
These values belong to the transition region between the dashed lines in
Fig. 3(a); (b) Same as (a), but for $N=10000$. The values of $\Lambda$
belong to the transition region between the dashed lines in Fig. 3(b). }
\end{figure}

\begin{figure} 
\centerline{
\epsfig{file=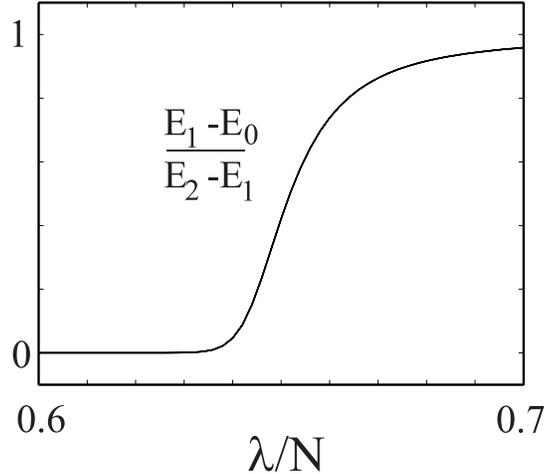,height=6.3 cm}}
\caption{The energy difference of the first excited state
and the ground state, and the energy difference of the second and first
excited states $(E_1-E_0)/(E_2-E_1)$, as a function of $\Lambda$ for
$N=1000$ and $U_1=3U_0$, calculated for the complete quantum field
theoretical model. The parameters are: $a^{\rm sc}_A=a^{\rm 
sc}_B=50$
nm, $a^{\rm sc}_{AB}=150$ nm, $x_0=3$ $\mu$m, $\hbar\omega= 100$ Hz
}
\end{figure}

\begin{figure} 
\centerline{
\epsfig{file=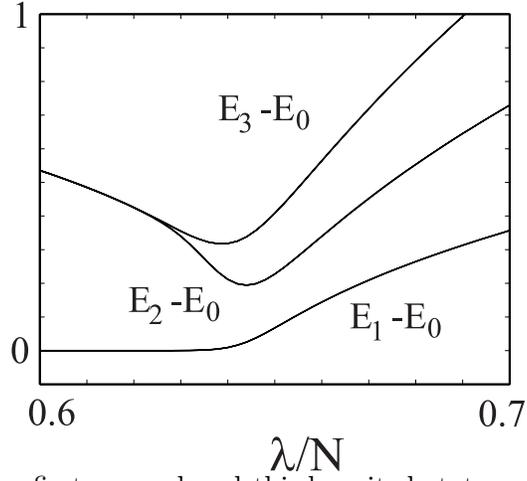,height=6.3 cm}}
\caption{The energy of the first, second and third
excited states with respect to the energy of the ground state as a
function of $\Lambda$ calculated for the complete quantum field
theoretical model. The parameters are the same as in Fig. 5. }
\end{figure}

\end{document}